%% file: paperII_2022.tex
\documentclass{article}
\usepackage{nips2003e,times}
\usepackage{latexsym,amsmath,amsfonts}
\usepackage{graphicx}

\title{The Lack of Convexity of the Relevance-Compression Function}

\author{
Albert E. Parker \\
Center for Biofilm Engineering\\
Department of Mathematical Sciences\\
Montana State University\\
Bozeman, MT USA\\
\texttt{parker@math.montana.edu} \\
\And
Tom\'{a}\v{s} Gedeon\\
Department of Mathematical Sciences\\
Montana State University\\
Bozeman, MT USA\\
\texttt{gedeon@math.montana.edu} \\
\And
Alexander G. Dimitrov\\
Department of Mathematics\\
Department of Neuroscience\\
Washington State University\\
Vancouver, WA\\
\texttt{alex.dimitrov@wsu.edu} \\
 }

\input{heading.tex}
\def\RR{{\rm I\kern -1.6pt{\rm R}}}
\def\Int{{\rm Int}}

\begin{document}

\maketitle

\begin{abstract}
\small{In this paper we investigate the convexity of the
  relevance-compression function for the Information Bottleneck and
  the Information Distortion problems. This curve is an analog of
  the rate-distortion curve, which is convex.
In the problems we discuss in this paper, the distortion function is
  not a linear function of the quantizer, and the relevance-compression
  function is not necessarily convex (concave), but can change its
  convexity.
We relate this phenomena with existence of first order phase
  transitions in the corresponding Lagrangian  as a function of the  annealing
  parameter.}
\end{abstract}

\section{Introduction}
In previous  work \cite{JDE, IEEEParker2010, Entropy2012} we
have described the bifurcation structure for solutions to problems
of the form
    \begin{eqnarray}
    \label{genmaxprob}
    \ba{c}
    \max_{q\in \Delta} G(q)\\
    D(q)\ge D_0
    \ea
    \end{eqnarray}
where $\Delta$ is a constraint space of valid conditional
probabilities, $G$ and $D$ are continuous, real valued functions
of $q$, smooth in the interior of $\Delta$, and the functions
$G(q)$ and $D(q)$ are invariant under the group of symmetries
$S_N$. This type of problem, which arises in Rate Distortion
Theory \cite{CoverInfo,GrayInfo} and Deterministic Annealing
\cite{RoseDAall}, is $NP$ complete \cite{NPcomplete} when $D(q)$
is the mutual information $I(X;Y_N)$ as in the Information
Bottleneck \cite{TishbyAgg,Infobneck,Noamthesis} and the
Information Distortion \cite{ADcoding,ADDI,CAMQ} methods. In this
paper, we address the relationship between the bifurcation
structure of solutions to (\ref{genmaxprob}) and the relevance
compression function \cite{Noamthesis}.

\section{Preliminaries}
\subsection{Rate Distortion Theory}
We assume that the random variable $X\in\{x_1,x_2,...,x_K\}$ is an
input source, and that $Y\in\{y_1,y_2,...,y_K\}$ is an output
source.  In rate distortion theory \cite{CoverInfo}, the random
variable $Y$ is represented by using $N$ symbols or {\em classes},
which we call $Y_N$, where we assume without loss of generality
that $Y_N\in\{1,2,...,N\}$.    We denote a stochastic clustering
or {\em quantization}, of the realizations of $Y$ to the classes
of $Y_N$, by $q(Y_N|Y)$.   To find the quantization that yields
the minimum information rate $I(Y;Y_N)$ at a given distortion, one
can find points on the rate distortion curve for each value of
$D_0\in[0,D_{\max}]$.   The rate distortion curve is defined as
\cite{CoverInfo,GrayInfo}
    \begin{eqnarray}
    \label{RDgray}%
    R(D_0):=
    \begin{array}{c}
    \min_{q\in \Delta} I(Y;Y_N)\\
    D(Y,Y_N) \le D_0
    \end{array},
    \end{eqnarray}
where $D(q)$ is a {\em distortion function}.    A quantization
$q(Y_N|Y)$ that satisfies (\ref{RDgray}) yields an approximation
of the probabilistic relationship, $p(X,Y)$, between $X$ and $Y$
\cite{ADcoding,TishbyAgg,Infobneck}. The constraint space $\Delta$
is the space of valid finite conditional probabilities $q(Y_N|Y)$,
where we will write $q(Y_N=\nu | Y=y_k)=\qnuk$.

The Information Bottleneck method
\cite{TishbyAgg,Infobneck,Noamthesis} uses the {\em information
distortion function} $D(q):=I(X;Y)-I(X;Y_N)$.   Since the spaces
$X$ and $Y$ are fixed, then $I(X;Y)$ is fixed, and so the rate
distortion problem (\ref{RDgray}) in the case of the Information
Bottleneck problem can rewritten as
    \begin{eqnarray}
    \label{tishby}
    R_I(I_0):=
    \ba{c}
    \max_{q\in \Delta} -I(Y;Y_N)\\
    I(X;Y_N)\ge I_0
    \ea
    \end{eqnarray}
where $I_0>0$ is some information rate. The function $R_I(I_0)$ is
referred to as the {\em relevance-compression function}
in~\cite{Noamthesis}.  Observe that there is a one to one
correspondence between $I_0$ and $D_0$ via $I_0=I(X;Y)-D_0$. To
solve the neural coding problem, the Information Distortion method
\cite{ADcoding,ADDI,CAMQ} considers a problem of the form
    \begin{eqnarray}
    \label{us}
    R_H(I_0):=
    \ba{c}
    \max_{q\in \Delta} H(Y_N|Y)\\
    I(X;Y_N)\ge I_0
    \ea
    \end{eqnarray}
where $H(Y_N|Y)$ is the conditional entropy.

\subsection{Annealing}
Using the method of Lagrange multipliers, an arbitrary problem of
the form (\ref{genmaxprob}) is rewritten as
    \begin{eqnarray}
    \label{genanneal}
    \max_{q\in \Delta} \left(G(q)) + \beta (D(q) - D_0)\right).
    \end{eqnarray}
As we will see, solutions of (\ref{genmaxprob}) are not always
solutions of (\ref{genanneal}).   Similarly, the problem
(\ref{tishby}) can be rewritten
\cite{TishbyAgg,Infobneck,Noamthesis} as
    \begin{eqnarray}
    \label{tishbyanneal}
    \max_{q\in \Delta} \left(-I(Y_N,Y) + \beta I(X;Y_N)\right),
    \end{eqnarray}
and  problem (\ref{us}) can be rewritten~\cite{ADcoding,ADDI,CAMQ}, in
analogy with Deterministic Annealing \cite{RoseDAall}, as
    \begin{eqnarray}
    \label{usanneal}
    \max_{q\in \Delta} \left(H(Y_N|Y) +\beta
    I(X;Y_N)\right).
    \end{eqnarray}
In (\ref{genanneal}), (\ref{tishbyanneal}) and  (\ref{usanneal}),
the Lagrange multiplier $\beta$ can be viewed as an annealing
parameter.

\subsection{Bifurcation Structure of solutions}
In \cite{parkerNIPS2002}, we presented an algorithm which
can be used to determine the bifurcation structure of stationary
points of (\ref{genanneal}) for each value of
$\beta\in[0,\beta_{\max})$ for some $\beta_{\max}>0$. These
stationary points are quantizers $q^*(\beta)\in\Re^{NK}$ where
there exists a vector of Lagrange multipliers $\lambda^*\in\Re^K$
such that the gradient of the Lagrangian of (\ref{genmaxprob}) is
a vector of 0's,
    \bes
    \label{gradientis0}
    \nabla_{q,\lambda}(G(q^*) + \beta D(q^*) + \sum_{k=1}^K\lambda^*_k(\sum_{\nu} q^*(\nu|y_k)
    -1)=\Z.
    \ees
This condition is also known as the Karush-Kuhn-Tucker necessary
condition for constrained optimality \cite{Nocedal}.   It is well
known in optimization theory that a stationary point, i.e. the
point satisfying (\ref{gradientis0}),  is a solution of
(\ref{genmaxprob}) if the matrix of second derivatives, the
Hessian $\Delta_q (G(q) + \beta D(q))$, is negative definite on
the kernel of the Jacobian of the constraints \cite{Nocedal}. We
have the following results.

\begin{thm}
\label{thm:isitasolution} \cite{JDE} A stationary point $q^*$,
is a solution of (\ref{genanneal})
if $\Delta_q (G(q^*) + \beta D(q^*))$ is negative definite on
    $\ker\left(
    \ba{c}
    I_K~~I_K~~...~~I_K
    \ea
    \right)$.  A stationary point $q^*$
is a solution of (\ref{genmaxprob})
if $\Delta_q (G(q^*) + \beta D(q^*))$ is negative definite on
    $\ker\left(
    \ba{c}
    \nabla_q D(q^*)^T\\
    I_K~~I_K~~...~~I_K
    \ea
    \right)$.
\end{thm}

\begin{figure}
\begin{tabular}{cc}
    A & B\\
    \includegraphics[width= .475 \textwidth]{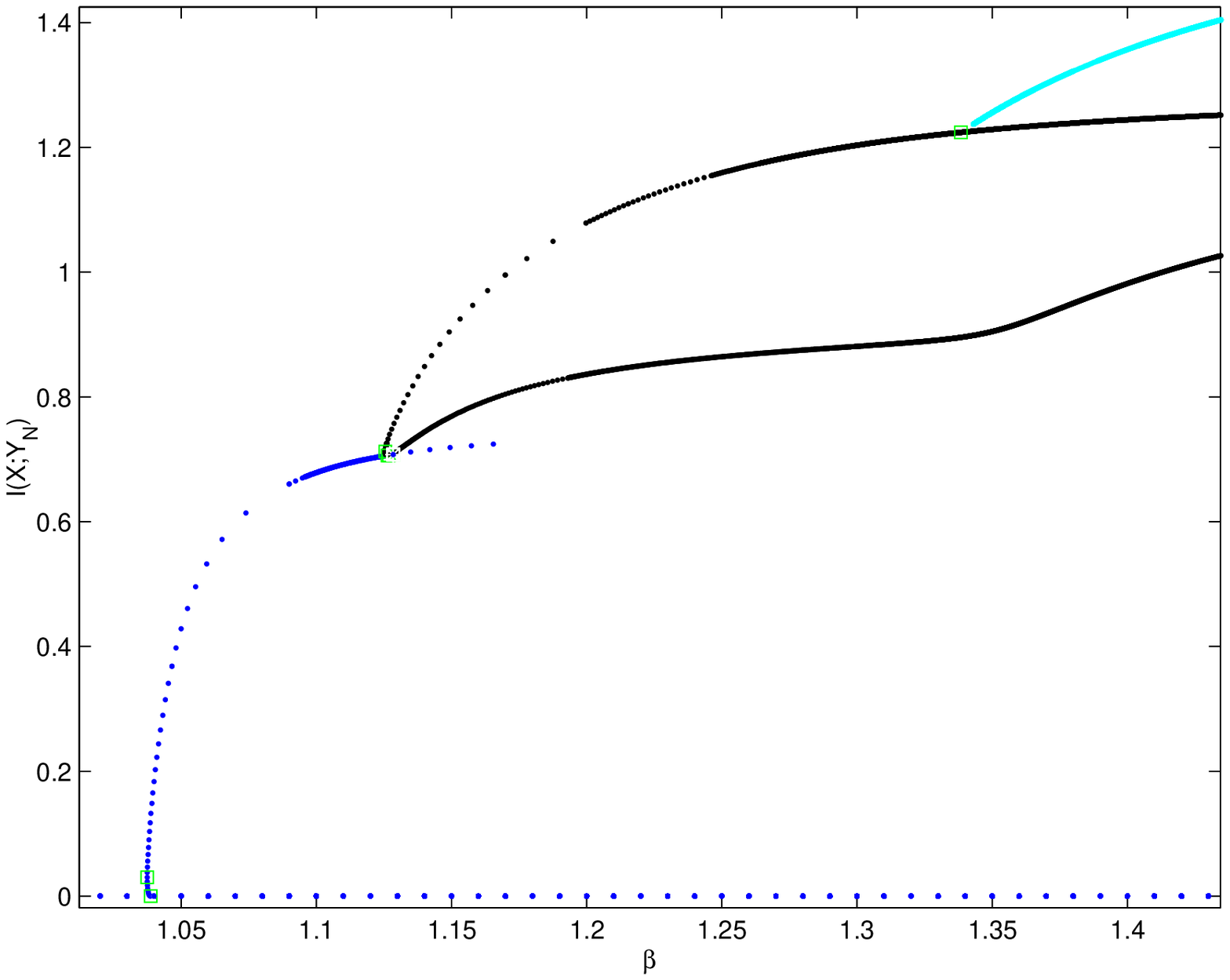}&
    \includegraphics[width= .475 \textwidth]{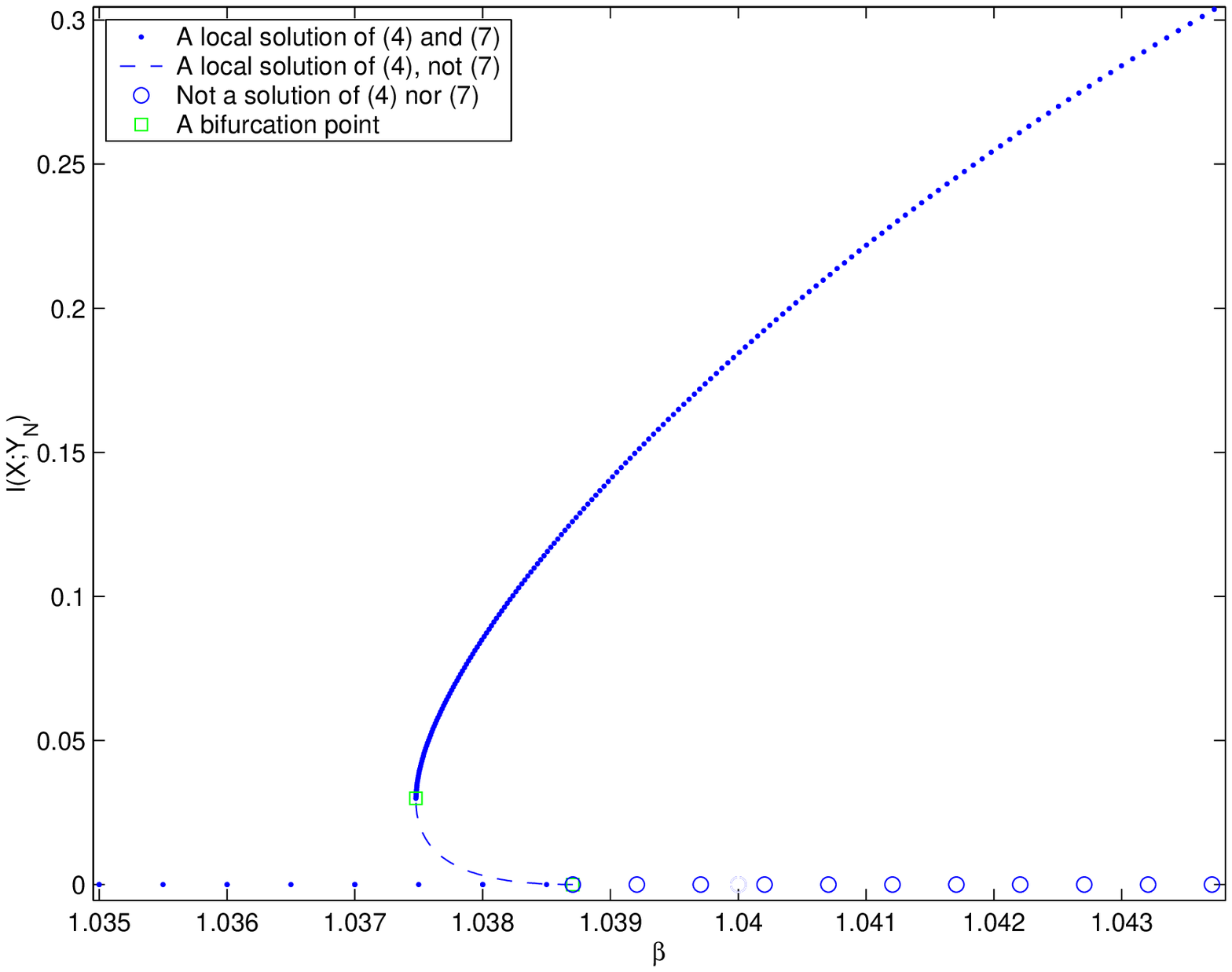}
\end{tabular}
\caption{(A) The bifurcation structure of stationary points of the
Information Distortion problem (\ref{us}), a problem of the form
(\ref{genmaxprob}).  We found these points by annealing in $\beta$
and finding stationary points for the problem (\ref{usanneal})
using the algorithm which we presented in \cite{parkerNIPS2002}. A
square indicates where a bifurcation occurs.  (B)  A close up of
the subcritical bifurcation at $\beta\approx 1.038706$, indicated
by an square.      Observe the subcritical bifurcating branch, and
the subsequent saddle-node bifurcation at $\beta\approx 1.037479$,
indicated by another square. We applied Theorem
\ref{thm:isitasolution} to show that the subcritical bifurcating
branch is composed of quantizers which are solutions of (\ref{us})
but not of (\ref{usanneal}).}
   \label{fig:bifstructure}
\end{figure}

From Theorem \ref{thm:isitasolution} we see that there may be
solutions of (\ref{genmaxprob}) which are not solutions of
(\ref{genanneal}).    We illustrate this fact  numerically. For
the Information Distortion problem (\ref{us})
\cite{ADcoding,ADDI,CAMQ}, and the synthetic data set composed of
a mixture of four Gaussians which the authors used in
\cite{ADcoding}, we determined the bifurcation structure of
solutions to (\ref{us}) by annealing in $\beta$ and finding the
corresponding stationary points to (\ref{usanneal})  (see Figure
\ref{fig:bifstructure}).

Similar to the results which we presented in
\cite{parkerNIPS2002}, the close up of the bifurcation at
$\beta\approx 1.038706$ in Figure \ref{fig:bifstructure}(B) shows
a subcritical bifurcating branch (a first order phase transition)
which consists of stationary points of the problem
(\ref{usanneal}).  By projecting the Hessian $\Delta_q (G(q^*) +
\beta D(q^*))$ onto each of the kernels referenced in Theorem
\ref{thm:isitasolution}, we determined that the points on this
subcritical branch are {\bf not} solutions of (\ref{usanneal}),
and yet they {\bf are} solutions of (\ref{genmaxprob}).

Furthermore, observe that Figure \ref{fig:bifstructure}(B)
indicates that a saddle-node bifurcation occurs at $\beta\approx
1.037479$. That this is indeed the case was proved
in~\cite{JDE}. In fact, for any problem of the form
(\ref{genanneal}), there are only two types of bifurcations to be
expected.

\begin{figure}
\begin{tabular}{cc}
A & B\\
    \includegraphics[width= .475 \textwidth]{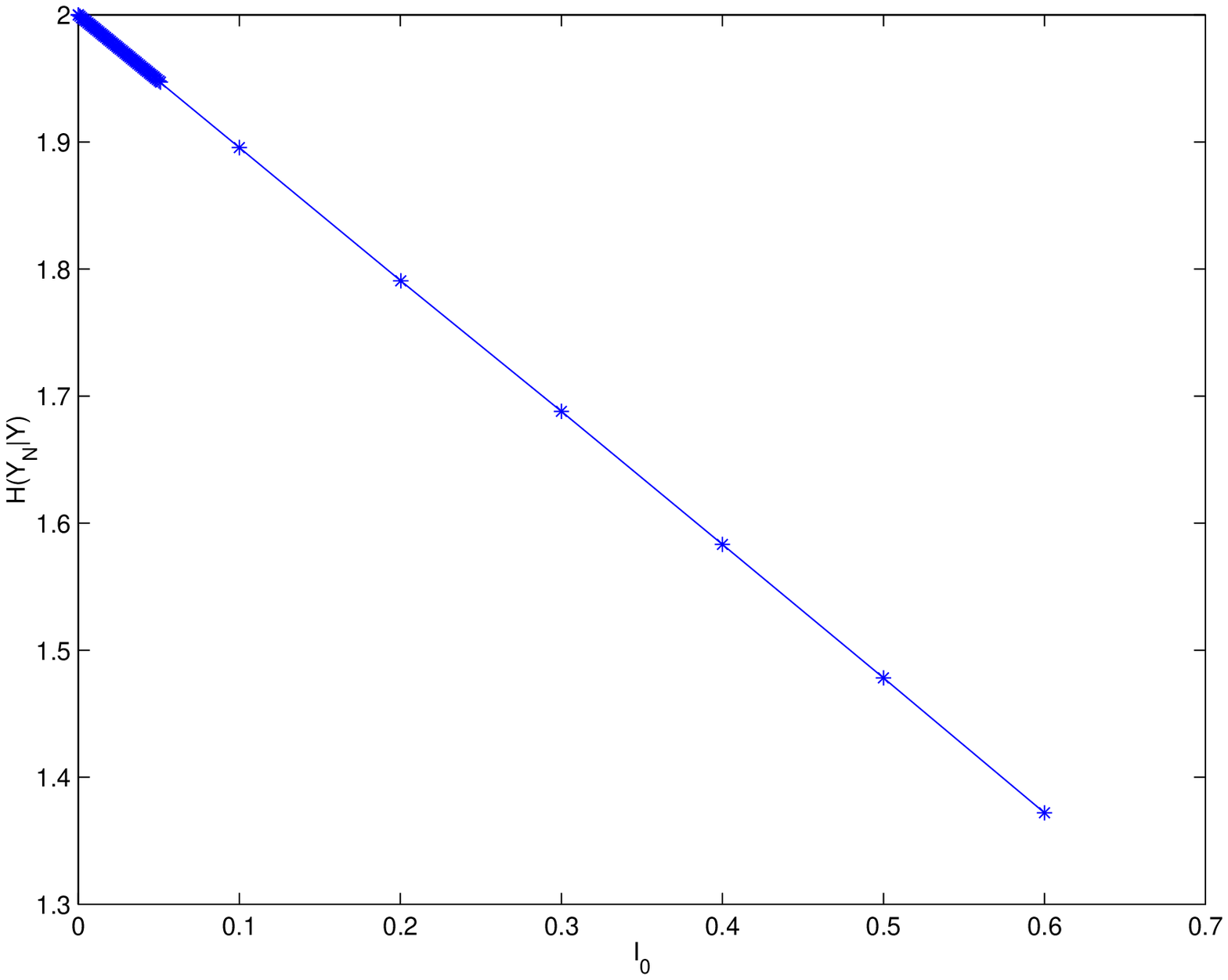} &
    \includegraphics[width= .475 \textwidth]{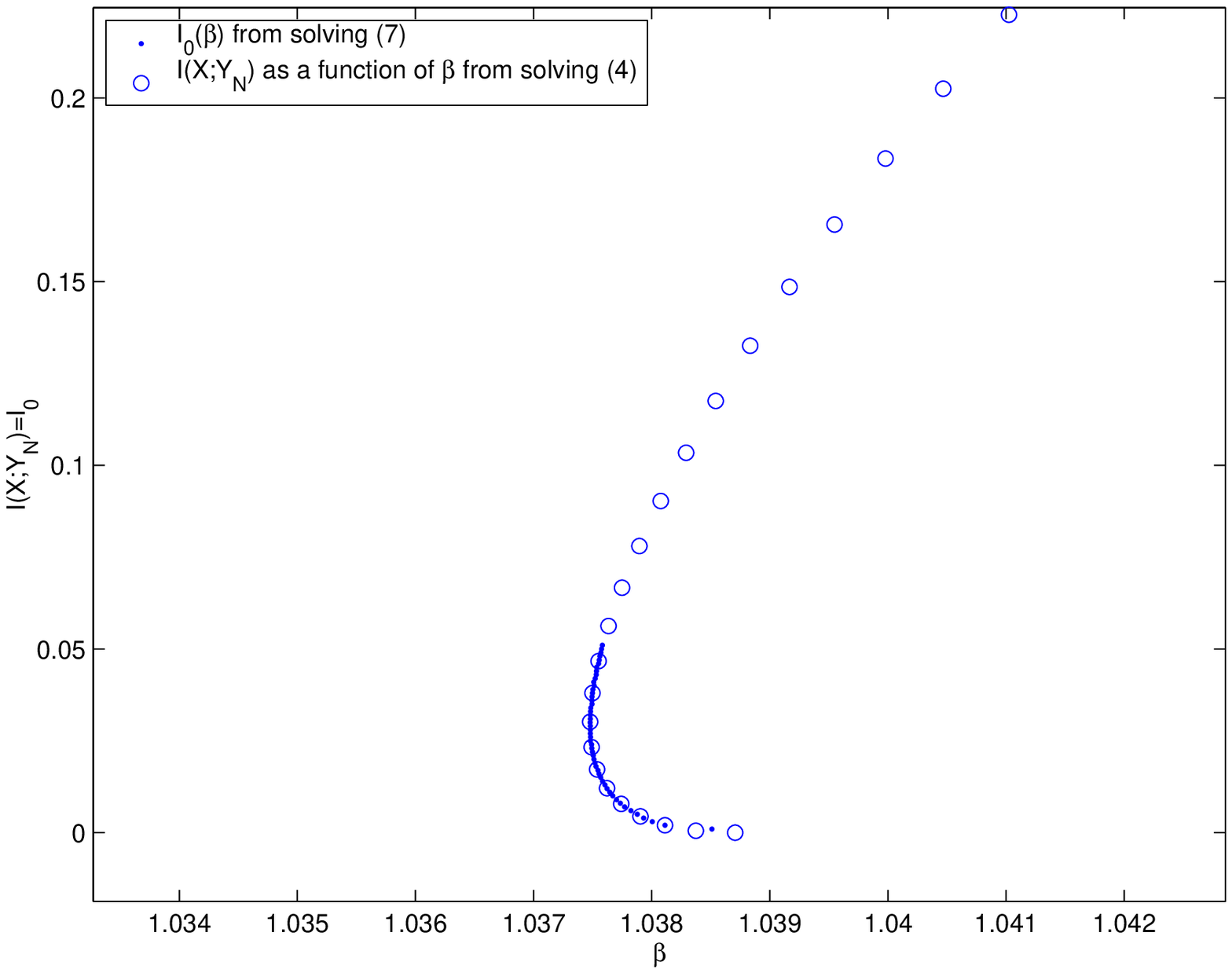}
\end{tabular}
\caption{(A)  The distortion curve $R_H(I_0)$ defined in
(\ref{usRD}). For each value of $I_0$, we solved the problem
(\ref{us}) to ascertain $H(Y_N|Y)$. Observe that the quantizers
which yield $I(X;Y_N)\ge I_0$ for the values $I_0\in[0,.03]$
correspond to solutions on the subcritical bifurcating branch
shown in Figure {fig:bifstructure}(B). (B)  For each value of
$I_0$, we solved (\ref{us}), and found the corresponding Lagrange
multiplier. This plot shows $I_0$ as function of this Lagrange
multiplier. This plot is identical to the subcritical bifurcation
shown in Figure \ref{fig:bifstructure}(B), which shows $I(X;Y_N)$
as a function of the annealing parameter $\beta$.}
   \label{fig:ratecurve}
\end{figure}

\begin{thm}
\cite{JDE} Generically, for problems of the form
(\ref{genmaxprob}), only symmetry breaking pitchfork-like and
saddle-node bifurcations occur.
\end{thm}
Clearly, the  existence of saddle-node bifurcation at $\beta\approx
 1.037479$ is tied to the  existence of
subcritical bifurcation (first order phase transition) at
$\beta\approx 1.038706$. We now investigate the connection between
existence of subcritical bifurcations and the convexity of the
relevance-compression function.

\section{The Relevance-Compression Function}
Given the generic existence of subcritical pitchfork-like and
saddle-node bifurcations of solutions to problems of the form
(\ref{genmaxprob}), a natural question arises:    What are the
implications for the rate distortion curve (\ref{RDgray})?    We
examine this question for the information distortion
$D(q)=I(X;Y)-I(X;Y_N)$, used by the Information Bottleneck and the
Information Distortion methods. Recall that the relevance-distortion
function is
   \begin{equation}
   \label{tishbyRD}
   R_I(I_0):= \max_{\Delta\cap {\cal Q}_{I_0}} -I(Y_N,Y).
   \end{equation}
where
    \[ {\cal Q}_{I_0} :=\{ q\in\Delta\:|\: I(X,Y_N) \geq I_0\}. \]
For
the Information Distortion problem the relevance-distortion function
is
    \begin{equation}\label{usRD}
    R_H(I_0):= \max_{\Delta\cap {\cal Q}_{I_0}} H(Y_N|Y).
    \end{equation}
In Figure \ref{fig:ratecurve}(A), we present a plot of $R_H(I_0)$,
which was computed using the same data set of a mixture of four
Gaussians which we used in Figures \ref{fig:bifstructure}(A) and
(B).    The plot was obtained by solving the problem (\ref{us})
for each value of $I_0$.

To make explicit the relationship between the bifurcation
structure shown in Figure \ref{fig:bifstructure}, which was
obtained by annealing in $\beta$, and the distortion curve shown
in Figure \ref{fig:ratecurve}(A), which was obtained by annealing
in $I_0$, we present Figure \ref{fig:ratecurve}(B).    When
solving (\ref{us}) for each $I_0$, we computed the corresponding
Lagrange multiplier $\beta$.   Thus, $\beta=\beta(I_0)$, which is
the curve we show in Figure \ref{fig:ratecurve}(B).  This plot is
matches precisely the subcritical bifurcating branch from Figure
\ref{fig:bifstructure}(B), which we obtained by annealing in
$\beta$.

\begin{lem}
For a fixed value of $I_0 >0$ the solution $q^*$ of (\ref{tishby})
and (\ref{us}) satisfies  the equality constraint $I = I_0$.
\end{lem}
\proof Assume  $q^*$ is a maximizer of (\ref{us}) and $I(q^*)
> I_0$. Then the constraint is not active and  we must have $\nabla
H(q^*) = 0$. Since  $\nabla
H(q^*) = 0$ implies $q^* = 1/N$, we get $I(q^*) = 0$. This is a
contradiction and thus  $I(q^*) = I_0$.

Now assume  $q^*$ is a maximizer of (\ref{tishby}) and $I(q^*)
> I_0$. Then again the constraint is not active and  we must have $\nabla
I(Y,Y_N) = 0$. Short computation shows that the condition $\nabla
I(Y,Y_N) = 0$ implies $q^* =(q_{\nu k})$ is of the form $q_{\nu k}
= a_\nu$, i.e. $q_{\nu k}$ does not depend on $k$. However, at
such value of $q^*$  we get $I(q^*) = 0$. This is again a
contradiction and thus  $I(q^*) = I_0$. \hfill $\Box$

As a consequence of the Lemma, for each $I_0 >0$ there exists a
Lagrange multiplier $\beta(I_0)$. The existence of subcritical
bifurcation branch implies that along this branch $\beta(I_0)$ is
not a one-to-one function of $I_0$, and therefore not invertible.


\subsection{Properties of relevance-compression function}
It is well known that if the distortion function $D(q)$ is linear
in $q$, that $R(D_0)$ is a non-increasing and convex
\cite{CoverInfo,GrayInfo}.  The proof of this result first
establishes that the rate-distortion curve is monotone and that it
is convex. These two properties together imply continuity and
strict monotonicity of the rate distortion curve. Since the
information distortion $D=I(X;Y)-I(X;Y_N)$ is not a linear
function of $q$, the convexity proof given in
\cite{CoverInfo,GrayInfo} does not generalize to prove that either
(\ref{tishbyRD}) or (\ref{usRD}) is convex.  Therefore we need to
prove continuity of the relevance-compression function using other
means.
\begin{lem}
The curves $R_H$ and $R_I$ are non-increasing curves on $I_0\in
[0,I_{\max}]$ and are continuous for $I_0 \in (0,I_{\max})$.
\end{lem}
\proof Observe that since $I(X,Y_N)$ is convex \cite{ADcoding} in
quantizer $q$, we have that
\[ {\cal Q}_{I_1} \subset {\cal Q}_{I_2} \mbox{ whenever } I_1 \geq
I_2 .\] Therefore, if $ I_1 \geq I_2$, then the maximization at
$I_1$ happens over a smaller set than in $\cal Q_{I_2}$, and so
$R(I_1) \leq R(I_2)$.

Now we prove continuity. Take an arbitrary $I_0 \in (0,I_{\max})$.
Let
    \[ M_{I_0}:= \{ y \:|\: y = H(q) \mbox{ where } q \in \Delta \cap
    {\cal Q}_{D_0} \} \]
be the range (in $\RR$) of the function $H(q)$ with the domain
$\Delta \cap {\cal Q}_{I_0}$. Given an arbitrary $\epsilon >0$,
let $M^\epsilon_{I_0}$ be an $\epsilon$ neighborhood of $M_{I_0}$
in $\RR$. A direct computation shows  that $\nabla_q H(q) = \Z$ if
and only if $q$ is homogeneous, i.e. $qnuk = 1/N$, where $N$ is
the number of classes of $Y_N$. Since $H(q)$ is continuous on
$\Delta$, then the set $ H^{-1}(M^\epsilon_{I_0})$ is a relatively
open set in $\Delta$. Because by definition $H(\Delta \cap {\cal
Q}_{I_0}) = M_{I_0}$, we see that
    \begin{equation}
    \label{incl}
    {\cal Q}_{I_0} \cap \Delta \subset H^{-1}( M^\epsilon_{I_0}) .
    \end{equation}
Furthermore, since $\nabla H(q)  \neq 0$ for $ q \in {\cal
Q}_{I_0}$, then, by the Inverse Mapping Theorem, $H^{-1}(
M^\epsilon_{I_0})$ is an open neighborhood of ${\cal
  Q}_{I_0}$.

The function $I(X;Y_N)$ is also continuous in the interior of
$\Delta$. Observe that
    \[ {\cal Q}_{I_0} = I^{-1}([I_0, I_{\max}]) \]
is closed, and thus ${\cal Q}_{I_0} \cap \Delta$ is closed and
hence compact. Thus, by  (\ref{incl}) $H^{-1}( M^\epsilon_{I_0})$
is an relatively open neighborhood of  a compact set ${\cal
Q}_{I_0} \cap \Delta$.  Therefore, since $I(X;Y_N)$ is continuous,
there exists a $\delta
>0$ such that  the set
\[ \Int{\cal Q}_{I_0 + \delta} \cap \Delta = I^{-1}((I_0+\delta,
D_{\max}])\cap \Delta \] is a relatively open set in $\Delta$ such
that \[{\cal Q}_{I_0} \cap \Delta \subset \Int{\cal Q}_{I_0 +
\delta}
 \subset H^{-1}( M^\epsilon_{I_0}) .\]
It then follows that
\[ |\max_{\Delta \cap {\cal Q}_{I_0+\delta}} H -  \max_{\Delta \cap {\cal
      Q}_{I_0}} H | <\epsilon. \]
By definition of the rate distortion function, this means that
    \[ |R(I) -R(I_0)|<\epsilon \mbox{ whenever } I-I_0 <\delta .\]
Since $\epsilon$ was arbitrary, this implies continuity of $R(I)$
at $I = I_0$. \hfill $\Box$

\subsection{The Derivative $\frac{\partial R}{\partial I_0}$}

In \cite{TishbyAgg,Noamthesis}, using variational notation, it is shown that
    $$\frac{\delta R_I}{\delta D} = -\beta.$$
For the sake of completeness, we will reprove this, acknowledging
explicitly the fact that the problems (\ref{tishby}) and
(\ref{us}) are constrained problems.
\begin{thm}\label{main}
If relevance-compression functions $R_I(I_0)$ and $R_H(I_0)$ are
differentiable, then
\begin{equation} \label{final}
\frac{dR}{dI_0} = -\beta(I_0) \mbox{ and } \frac{d^2R}{dI_0^2} = -\frac{d\beta(I_0)}{dI_0}
\end{equation}
\end{thm}
\begin{cor}
Since $\frac{d\beta(I_0)}{dI_0}$ changes sign at saddle-node
bifurcation, then the relevance-compression functions $R_I(I_0)$
and $R_H(I_0)$ are neither concave, nor convex.
\end{cor}
{\bf Proof of Theorem~\ref{main}:}
We start with
\begin{equation}
  \label{eq:prob}
  \max_{q\in\Delta} R(q) + \beta D(q) + \sum_k \lambda_k (\sum_\nu q_{\nu k}-1),
\end{equation}
where  $R(q)$ is one of $R_I(q):= H(Y_N|Y), R_H(q):=-I(Y_N,Y)$. We
parameterize the solution $q^*$ locally by $\beta$. This can be
done everywhere except if $q^*$ is at a saddle-node bifurcation.
At $q^*(\beta)$,
\begin{equation}
  \label{eq:soln}
  \nabla_q R + \beta \nabla_q D + \vec{\lambda}= 0.
\end{equation}
\begin{lem}
 For $q \in \Delta$,
    \begin{eqnarray}
  \label{eq:qdot}
  q \cdot \nabla_q R_H = R_H+1,~~q \cdot \nabla_q R_I = R_I,
  \mbox{ and }q \cdot \nabla_q I = I.
  \end{eqnarray}
\end{lem}
\proof
Direct calculation.
\hfill $\Box$

Hence  (\ref{eq:soln}) implies
\begin{equation}
  \label{eq:qdotres}
  R(q^*(\beta))+c+\beta I(q^*(\beta)) + q \cdot \vec{\lambda}= 0
\end{equation}
Here $c=1$ for $R_H$ and $c=0$ for $R_I$ is a constant. For $q \in
\Delta$, we set
\[ \Lambda(\beta):= q \cdot \vec{\lambda} = \sum_k \lambda_k .\]
  The equation (\ref{eq:qdotres})  defines a relation between $R$ and
 $I$. Recall, that we can always express $\beta= \beta(I_0)$. Then the
 term  $I(q^*(\beta(I_0))) = I_0$ and we have a relationship
\begin{equation}
  \label{eq:I}
  R(I_0)+c+\beta(I_0) I_0 + \Lambda(I_0)= 0.
\end{equation}
We differentiate (\ref{eq:I}):
\begin{eqnarray}
  \label{eq:dRdI}
  \frac{dR}{dI_0} + \frac{d\beta}{dI_0} I_0 + \beta(I_0) + \frac{d\Lambda}{dI_0} &=& 0 \Rightarrow
\end{eqnarray}
which shows that
$\frac{dR}{dI_0}  =  -\frac{d\beta}{dI_0}(I_0 +
\frac{d\Lambda}{d\beta}) - \beta(I_0)$ since
$\frac{d\Lambda}{dI_0}=\frac{d\Lambda}{d\beta}\frac{d\beta}{dI_0}$.

In \cite{ADcoding}(equation  (10))  we have an explicit expression
for $\lambda_k$ as a function of $\beta$:
\begin{equation}
  \label{eq:lk}
  \lambda_k = p_k(1-\ln \sum_\nu e^{ \beta(\nabla I)_{\nu k}/p_k}).
\end{equation}
Differentiating this with respect to $\beta$ yields
\[\frac{d\lambda_k}{d\beta} = -p_k\frac{\sum_\nu e^{ \beta(\nabla
I)_{\nu k}/p_k} (\nabla I)_{\nu k}/p_k}{\sum_\nu e^{ \beta(\nabla
I)_{\nu k}/p_k}}.\] Since $\Lambda = \sum_k \lambda_k$, this
implies that
    \bes
  \frac{d \Lambda}{d\beta} &=& \sum_k \frac{d\lambda_k}{d\beta}  =  -\sum_{\nu k}\frac{e^{ \beta(\nabla I)_{\nu k}/p_k} }{\sum_\mu
  e^{ \beta(\nabla I)_{\mu k}/p_k}} (\nabla I)_{\nu k}
  \ees
For a solution $q^*$, $\frac{e^{ \beta(\nabla I)_{\nu k}/p_k} }{\sum_\mu e^{ \beta(\nabla I)_{\mu k}/p_k}} = q^*_{\nu k}$ (\cite{ADcoding}, (12)), hence
\[
  \frac{d \Lambda}{d\beta}|_{q^*} = -\sum_{\nu k} q^*_{\nu k} (\nabla I)_{\nu k} = -q^* \cdot \nabla I = -I \]
This shows that the term $I_0 + \frac{d \Lambda}{d\beta} = 0$ at
$q^*$, hence from (\ref{eq:dRdI}) we get the first part of
(\ref{final}). The second part follows immediately.   \hfill
$\Box$


\subsubsection*{Acknowledgments}
This research is partially supported by NSF grants DGE 9972824,
MRI 9871191, and EIA-0129895; and NIH Grant R01 MH57179.

\bibliographystyle{unsrt}
\bibliography{../../refs/parker,../../refs/kerL,../../refs/alex,../../refs/coding,../../refs/filters,../../refs/JPMproposal,../../refs/lit}

\end{document}

%% file: heading.tex
\newtheorem{thm}{Theorem}

\newtheorem{cor}[thm]{Corollary}

\newtheorem{lem}[thm]{Lemma}

\newcommand{\proof}{\noindent{\em Proof.} \hspace{.2in}}

\newcommand{\Int}{{\rm Int}}

\newcommand{\ba}{\begin{array}}
\newcommand{\ea}{\end{array}}
\newcommand{\be}{\begin{eqnarray}}
\newcommand{\ee}{\end{eqnarray}}
\newcommand{\bes}{\begin{eqnarray*}}
\newcommand{\ees}{\end{eqnarray*}}

\def\qnuk{q_{\nu k}}

\def\ln{{\rm ln}}

\def\Z{\pmb 0}